\documentclass[sigconf,10pt]{acmart}
\usepackage{enumitem}
\usepackage{graphicx}
\graphicspath{ {./images/} }
\usepackage{soul}
\AtBeginDocument{%
  \providecommand\BibTeX{{%
    \normalfont B\kern-0.5em{\scshape i\kern-0.25em b}\kern-0.8em\TeX}}}

\acmYear{2022}\copyrightyear{2022}
\setcopyright{rightsretained}
\acmConference[ACM MobiCom '22]{The 28th Annual International Conference on Mobile Computing and Networking}{October 17--21, 2022}{Sydney, NSW, Australia}
\acmBooktitle{The 28th Annual International Conference on Mobile Computing and Networking (ACM MobiCom '22), October 17--21, 2022, Sydney, NSW, Australia}
\acmPrice{15.00}
\acmDOI{10.1145/3495243.3558759}
\acmISBN{978-1-4503-9181-8/22/10}

\begin{document}

\title{IMAP: Individual huMAn mobility Patterns visualizing platform}

\author{Yisheng Alison Zheng}
\affiliation{%
  \institution{The University of Sydney}
  \city{Sydney }
  \country{Australia}}
\email{yzhe5242@uni.sydney.edu.au}
\orcid{0000-0003-1192-6434}

\author{Amani Abusafia}
\affiliation{%
  \institution{The University of Sydney}
  \city{Sydney }
  \country{Australia}}
\email{amani.abusafia@sydney.edu.au}
\orcid{0000-0001-9159-6214}

\author{Abdallah Lakhdari}
\affiliation{%
  \institution{The University of Sydney}
  \city{Sydney }
  \country{Australia}}
\email{abdallah.lakhdari@sydney.edu.au}
\orcid{0000-0001-8005-1534}

\author{Shing Tai Tony Lui}
\affiliation{%
  \institution{The University of Sydney}
  \city{Sydney }
  \country{Australia}}
\email{slui2950@uni.sydney.edu.au}
\orcid{0000-0001-9698-4697}

\author{Athman Bouguettaya}
\affiliation{%
  \institution{The University of Sydney}
  \city{Sydney }
  \country{Australia}}
\email{athman.bouguettaya@sydney.edu.au}
\orcid{0000-0003-1254-8092}

\renewcommand{\shortauthors}{Yisheng and Amani,  et al.}

\begin{abstract}
Understanding human mobility is essential for the development of smart cities and social behavior research. Human mobility models may be used in numerous applications, including pandemic control, urban planning, and traffic management. The existing models' accuracy in predicting users' mobility patterns is less than 25\%. The low accuracy may be justified by the \textit{flexible nature} of human movement. Indeed,  humans are not rigid in their daily movement. In addition, the rigid mobility models may result in missing the hidden regularities in users' records. Thus, we propose a novel perspective to study and analyze human mobility patterns and capture their flexibility. Typically, the mobility patterns are represented by a sequence of locations. We propose to define the mobility patterns by abstracting these locations into a set of places. Labeling these locations will allow us to detect \textit{close-to-reality} hidden patterns. We present \textit{IMAP}, an Individual huMAn mobility Patterns visualizing platform. Our platform enables users to \textit{visualize a graph} of the places they visited based on their history records. In addition, our platform displays the most \textit{frequent mobility patterns} computed using a modified PrefixSpan approach.

\end{abstract}


\maketitle
\section{Introduction}

\textit{Human Mobility patterns} refers to the sequences of frequently visited locations by a user \cite{gonzalez2008understanding}. Detecting the mobility patterns of individuals is essential in a wide range of applications, including pandemic prevention \cite{eubank2004modelling,wang2019urban}, urban planning \cite{xia2018exploring,Amani2022QoE} crowd management \cite{solmaz2019toward,zhou2020understanding,lakhdari2021fairness}, and  location-based services \cite{karamshuk2011human,lakhdari2016link,yao2022wireless}. Several research studies have found that human mobility is highly predictable \cite{gonzalez2008understanding,song2010limits}. This predictability can be related to the regularity of our daily routines \cite{yang2014modeling,zhou2018understanding,lakhdari2021proactive}. Acquiring the human mobility patterns mainly focuses on analyzing the spatio-temporal attributes and potential regularities hidden in individual, and population movement trajectories \cite{wang2019urban,solmaz2019toward}. Several studies proposed models to represent and predict the human mobility patterns \cite{haifeng2021human}. The availability of heterogeneous location-based data through social networks provides an unprecedented opportunity for a more in-depth exploration of human mobility patterns from a quantitative and microscopic perspective \cite{wang2019urban}.
 
Detecting individuals' mobility patterns requires mining their history of visited places. A mainstream of approaches that detect users' mobility patterns focus on using deep learning techniques to predict users' next possible location \cite{luca2021survey, haifeng2021human}. However, these approaches' accuracy is usually low and ranges between 8 - 25\%. Indeed, detecting a rigid mobility pattern for a user is not feasible due to the flexible nature of the human movement \cite{gonzalez2008understanding,wang2019urban,song2010limits}. Indeed, humans don't have rigid behavior in their daily movements. For example, let us assume a user has \textit{daily} lunch between 12:00 and 13:00  and enjoys Thai cuisine. However, they may go to a different Thai restaurant every day, e.g., Caysorn Thai restaurant on the first day, Seasoning Thai restaurant on the second day, and a Thai Pothong restaurant on the third day. Even though this user eats out regularly, it is challenging to detect that pattern since these venues have different locations. Therefore, we propose to label these locations by their place type to define the users' mobility patterns accurately.
For instance, given the previous example, if we label the three Thai restaurants as "Thai restaurant". A \textit{close-to-reality} hidden pattern will be detected.\looseness=-1

 
We propose to demonstrate the new approach of labeling users' locations by implementing a platform that visualizes users' records and patterns. The platform takes the users' records after labeling the locations and draws \textit{a graph of the places} they visited (see Fig.\ref{fig:user}). In addition, it visualise the detected \textit{frequent mobility patterns} using a modified PrefixSpan algorithm \cite{pei2004mining}.\looseness=-1


\vspace{-5pt}
\section{Overview}

Our platform allows users to visualize an individual user's mobility graph and patterns. Users may select the default dataset or upload their history of check-in places using a social network platform, e.g., Foursquare. In this demo, we used the Foursquare dataset as our default dataset \cite{yang2014modeling}. The Foursquare dataset is a \textit{geo-tagged social media} (GTSM) dataset where users check in to venues they have visited. We used the New York dataset, which has  227,428 check-in records. The platform involves three main steps (see Fig.\ref{fig:system}):

\begin{figure}[!t]
    \centering
    \includegraphics[width=\linewidth]{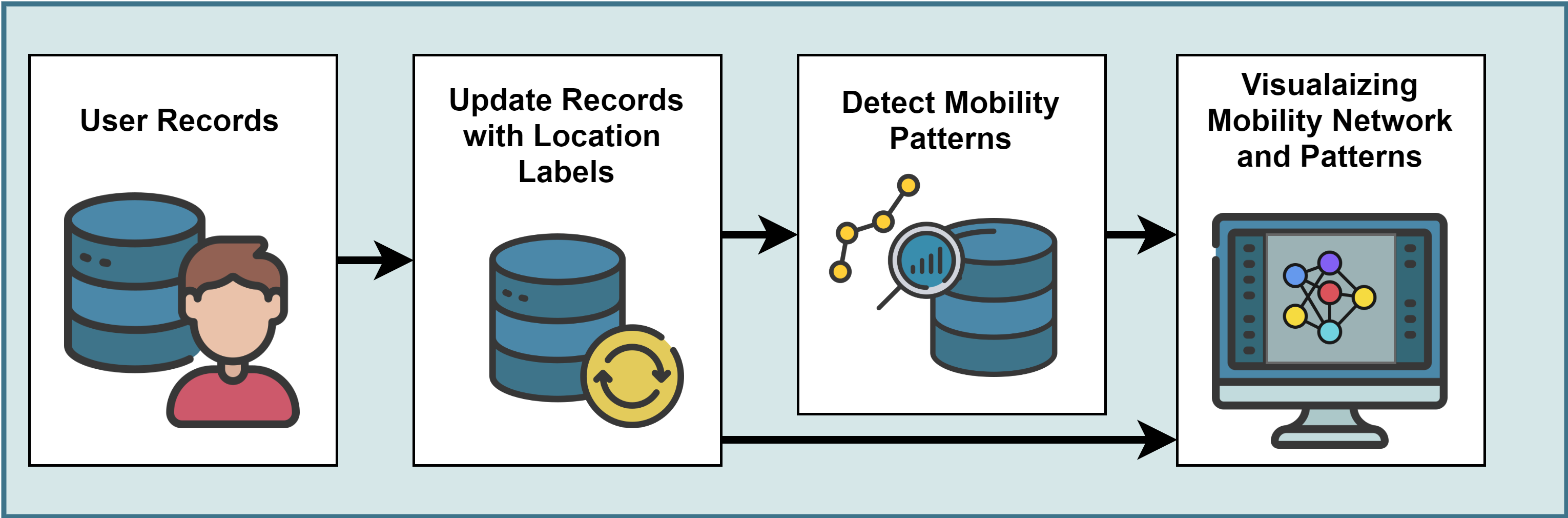}
    \caption{System overview}
    \label{fig:system}
\end{figure}

\begin{figure}[!t]
    \centering
    \includegraphics[width=\linewidth]{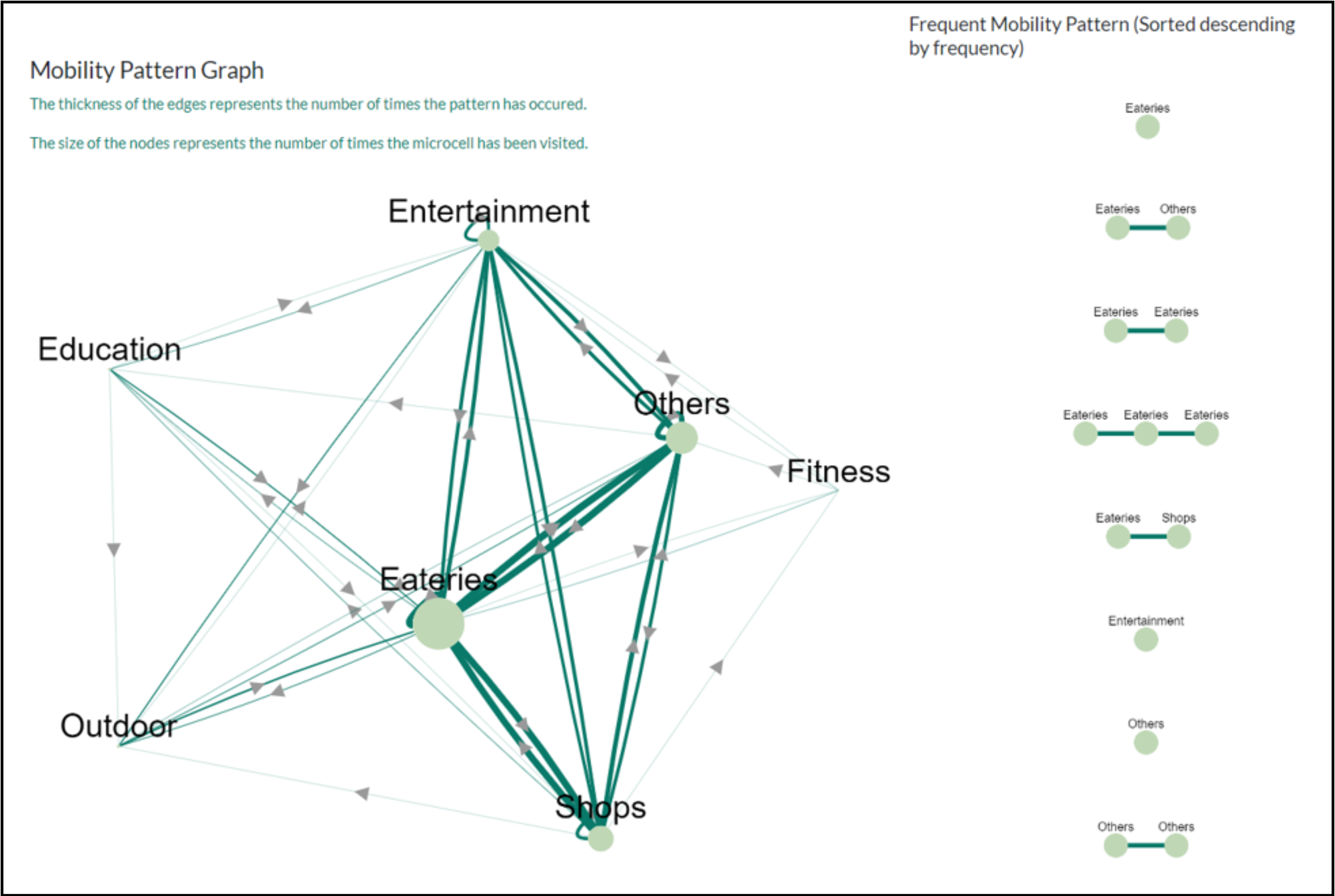}
    \caption{Example of a user mobility graph and patterns using \textit{IMAP} }
    \label{fig:user}
\end{figure}

\begin{enumerate} [leftmargin=15pt]
    \item \textbf{Update user records:} The user records of visited places will be relabeled by the name of the places. For instance, all Thai restaurants will be labeled as "Thai restaurant". The labeling rules can be decided by domain experts based on the purpose of the pattern mining.\looseness=-1
    \item \textbf{Detect mobility patterns:} After updating the users' rec-ords, a modified PrefixSpan algorithm is used to detect the frequent mobility patterns of the user \cite{pei2004mining}.
    \item \textbf{Visualize the mobility graph and patterns:} In this step, we visualize the user mobility graph and patterns using their relabelled records and mobility patterns (see Fig.\ref{fig:user}). The thickness of the edges represents the frequency of the pattern between two nodes. In Addition, the size of the nodes represents the times a place has been visited.
\end{enumerate}




\section{Demo Setup}
Our demonstration will exhibit the \textit{IMAP} web application features. Additionally, we will display a recorded video of the entire process of using the platform to present and interact with one of the default user's users records in real-time. The video can be found at this link: https://youtu.be/94bHbmBLXTs. To demonstrate the efficiency of our platform, we will also run our framework using the Foursquare public dataset on a laptop \cite{yang2014modeling}. Visitors of our booth can choose from a list of available users to visualize their network and their mobility patterns. They can also interact with the network by visualizing statistics about the displayed mobility patterns. If any of the audience is willing to share their history, we may upload it to the platform and visualize their patterns. 
\section*{Acknowledgment}
This research was partly made possible by  LE220100078 and LE180100158 grants from the Australian Research Council. The statements made herein are solely the responsibility of the authors. 
\bibliographystyle{unsrt} 
\bibliography{main}

\end{document}